\documentclass[twocolumn,showpacs,preprintnumbers,amsmath,amssymb,floatfix]{revtex4}
\input{epsf}

\usepackage{graphicx}% Include figure files
\usepackage{dcolumn}% Align table columns on decimal point
\usepackage{bm}% bold math

\begin{document}

\title{Single-atom aided probe of the decoherence of a Bose-Einstein condensate}
\author{H. T. Ng and S. Bose}
\affiliation{{Department of Physics and Astronomy, University
College London, Gower Street, London WC1E 6BT, United Kingdom}}
\date{\today}

\begin{abstract}
We study a two-level atom coupled to a Bose-Einstein condensate. We
show that the rules governing the decoherence of mesoscopic
superpositions involving different classical-like states of the
condensate can be probed using this system. This scheme is
applicable irrespective of whether the condensate is initially in a
coherent, thermal or more generally in any mixture of coherent
states. The effects of atom loss and finite temperature to the
decoherence can therefore be studied. We also discuss the various
noise sources causing the decoherence.
\end{abstract}


\maketitle
\section{Introduction}
Decoherence is a process of losing quantum superpositions due to
entanglement between the system and its environment \cite{Zurek}.
Studies of the decoherence are pivotal to understanding the
emergence of the classical world from an underlying quantum
substrate \cite{Zurek,Joos}. This is because if decoherence did not
suppress quantum superpositions in macroscopic systems, then one
would end with situations such as the Schr\"{o}dinger's cat
\cite{Schrodinger} which are not observed in practice. By now,
superpositions and decoherence of microscopic systems have been
observed in several experiments such as cavity QED
\cite{Brune,Raimond} and trapped ions \cite{Leifried,Myatt0}.
However in view of the fundamental implications for the quantum to
classical transition, it is necessary to gradually extend such
experimentation to superpositions in macroscopic systems, perhaps
tackling mesoscopic systems at first. In this context, several years
ago, Leggett and co-workers proposed the possibility of observing
superpositions of macroscopically distinct flux states in
superconducting systems \cite{garg-leggett,Leggett}, an idea which
has only recently been experimentally realized \cite{Rouse}. Another
class of experiments, clearly probing quantum superpositions and
their decoherence in the mesoscopic domain, is the diffraction of
large molecules \cite{Arndt,Hackermuller}. In order to move to
mesoscopic systems with slightly larger masses and investigating
their decoherence, there exist a number of proposals for using
nano-scale movable mirrors coupled to quantized light in cavities
\cite{Bose2}, or coupled to a Cooper pair box \cite{Armour} or to
photons in an interferometer \cite{Marshall}. The above three
schemes \cite{Bose2,Armour,Marshall} rely on the common idea of
coupling a microscopic (coherent) system to a mesoscopic or
macroscopic (decoherent) system to probe the decoherence of the
latter. In this paper, we formulate a scheme based on the same
general principle for a completely different mesoscopic system,
namely a Bose-Einstein condensate. While in the earlier work
involving nano-scale mirrors it is a superposition of spatially
separated coherent states whose decoherence is studied, in the
current paper it is the decoherence of a superposition of coherent
states with different phases (or relative phases in the case of two
mode condensates) which will be studied.

\begin{figure}[ht]
\includegraphics[height=4cm]{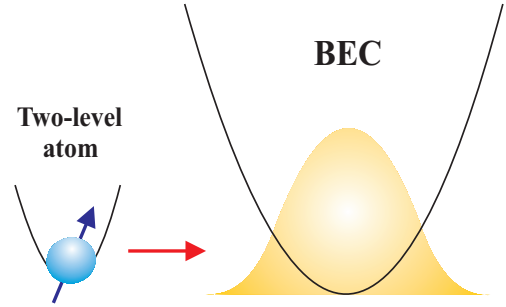}
\caption{ \label{fig1} (Color online) A single two-level atom is
coupled to a Bose-Einstein condensate via coherent collisions. The
single atom is trapped in a state-dependent potential such that the
upper state $|e\rangle$ interacts with the trapped condensate
merely. }
\end{figure}

Recently, a general scheme has been proposed \cite{Bose1} to probe
the decoherence of a mesoscopic harmonic oscillator with a qubit
whose state couples to the position of the oscillator. It provides a
simultaneous method to generate superpositions and probe the
decoherence of a mesoscopic system. The basic assumption in this
scheme is just to maintain the coherence of the qubit during the
whole period of the experiment, while the oscillator is allowed to
decohere (it is this decoherence that the scheme aims to detect).
When a qubit and the oscillator start in appropriate states, they
become entangled and a superposition in their joint system is
created. The special form of the coupling entails that after a
certain period of time evolution the qubit naturally disentangles
from the oscillator and the oscillator is brought back to its
original state. The evidence of the decoherence of the oscillator is
then imprinted on the partial coherence of the qubit. This enables
us to determine the decoherence rate of the mesoscopic oscillator by
measuring the qubit's state. Perhaps the strongest positive feature
of this type of scheme is the fact that the mesoscopic oscillator
will not have to be cooled to a pure or nearly pure state in order
for the scheme to be successful in probing its decoherence
\cite{Bose2,Armour,Marshall,Bose1}, in fact it can be in a thermal
state at arbitrary temperature. This happens because of the special
nature of the qubit-oscillator coupling and the fact that the
mesoscopic system is never directly probed. Naturally an important
question is whether such a scheme can be formulated for other kinds
of qubits (the qubits used earlier have been photon number/path
\cite{Bose2,Marshall} or superconducting \cite{Armour}) such as
atomic hyperfine levels which are more coherent, or oscillators
which are more coherent than nano-scale mirrors but at the same time
still mesoscopic is some sense. More fundamentally, would such a
scheme work, particularly would it still be applicable to thermal
states of the oscillator if the qubit was coupled to a different
variable of the oscillator instead of its position? The system
studied in this paper shows that the answer to both the above
questions is affirmative.

Atomic Bose-Einstein condensates (BEC's) are mesoscopic systems with
a low dissipation rate. They are promising candidates for observing
decoherence.  In fact, experimental studies have rapidly developed
since the Bose-Einstein condensation of atomic gases was observed in
a magnetic trap in 1995 \cite{Bradley}.  For example, the quantum
phase transition from a superfluid phase to a Mott insulator phase
has been observed using atoms in optical lattices \cite{Greiner}.
The trapping of multicomponent BEC's of ${}^{87}$Rb \cite{Myatt}
have been realized.  The control of scattering lengths using
Feshbach resonances have also been reported
\cite{Inoyue,Cornish,Marte,Roati}. These demonstrate the
sophisticated techniques available to manipulate ultracold atoms
precisely. Notably, schemes for deterministic single atom
preparation in the ground state in a trap potential have been
proposed \cite{Diener,Mohring}. A single-atom preparation and its
detection is expected to be realized in the near future
\cite{Treutlein}. This paves the way for studying the dynamics of a
single atom coupled to a superfluid BEC \cite{Recati}. In
particular, schemes for cooling a single atom \cite{Daley} and
probing the phase fluctuations of BEC \cite{Bruderer} have been
suggested with precisely such an atom-BEC coupled system. In this
paper, we will present a foundational application of such a system
in probing the decoherence of a BEC.

Naturally, because of the low dissipation rates of BECs, there have
been quite a few proposals for preparing them directly in
non-classical states such as Schr\"{o}dinger cat states
\cite{Scatbec1,Scatbec2,Scatbec3} even without an additional system
such as a single atom. If realized, these proposals will also enable
one to detect the decoherence that such states of a BEC experience.
However, the non-classical states of these papers, being
superpositions of states differing by large atom numbers in a given
mode, are highly decoherent and their preparation is extremely hard
(may even require one to engineer the environment \cite{Scatbec2}
and may be destroyed even due to scattering a single photon
\cite{Scatbec3}). Moreover to prepare such states one needs to have
a BEC in its ground state and thus whether the detection of
decoherence will work for a thermal state is unclear. Most
importantly, it is a superposition involving the most
``classical-like" states of a harmonic oscillator, namely its
coherent states, whose decoherence is the most relevant for studying
the transition from the quantum world to the classical world (as
among all the available quantum states of a harmonic oscillator, the
coherent states are the closest to classical states because they
have equal and minimum uncertainty in all variables). The scheme of
this paper is more ideally placed to detecting decoherence of
superpositions involving distinct coherent states as opposed to the
Schr\"{o}dinger cat states of mode occupation number
\cite{Scatbec1,Scatbec2,Scatbec3}.

In this paper, we study a two-level atom coupled to an atomic BEC
via coherent collisions.  A trapped BEC behaves like a harmonic
oscillator \cite{Milburn1,Dunningham,Ng} such that we can attempt to
formulate a scheme similar to Refs.\cite{Bose1,Bose2} to probe the
decoherence of a BEC with a single atom.  In this way, the aim is to
test the decoherence  of superpositions of a mesoscopic oscillator.
The heuristic formula for the decoherce rate of a superposition of
two coherent states is given by $(\hbar=1)$ \cite{Bose1,Bose2}
\begin{eqnarray}
\tilde{\Gamma}=2\Gamma\Big(\bar{n}+\frac{1}{2}\Big)D^2,
\end{eqnarray}
where $\Gamma$ is the characteristic damping constant, $D$ is the
separation of two coherence states in phase space,
$\bar{n}=[\exp(\omega/k_BT)-1]^{-1}$, $\omega$ is the frequency of
the oscillator, $k_B$ is the Boltzmann constant and $T$ is the
temperature.

We will show that our scheme can be applied to detect the
decoherence rate of an initial coherent state and even a thermal
mixed state. Finite temperature effects on the decoherence of a BEC
can thus be studied. Particle loss of the condensates gives rise to
the decoherence \cite{Davidovich}. Atom loss is caused by inelastic
collisions between atoms, and it is the dominant source of
decoherence in the BEC's \cite{Burt,Stenger}.

This paper is organized as follows:  In Sec II, we introduce the
coupled atom-BEC system.  The coupling of a single two-level atom to
a single BEC and a two-component BEC are examined.   Both cases can
be shown to map to a qubit-oscillator model.  In Sec III, we
describe our scheme to probe the decoherence of condensates when
they are initially in coherent and thermal states. In Sec IV, we
discuss the main decoherence sources of the BEC's and the
limitations of tuning the scattering length with Feshbach resonances
in Sec V.

\section{System}
We consider a single atom with two hyperfine spin states trapped in
the motional ground state of a potential.  The Hamiltonian of this
two-level atom is written as
\begin{eqnarray}
H_{s}&=&\omega_0(|e\rangle\langle{e}|-|g\rangle\langle{g}|),
\end{eqnarray}
where $\omega_0$ is the energy splitting, $|e\rangle$ and
$|g\rangle$ are the upper and lower states respectively.  This
spin-half system can be expressed in terms of Pauli operator:
$\sigma_z=|e\rangle\langle{e}|-|g\rangle\langle{g}|$,
$\sigma_+=|e\rangle\langle{g}|$ and $\sigma_-=|g\rangle\langle{e}|$.
Thus, the Hamiltonian can be cast in the form as
\begin{eqnarray}
H_{s}&=&\omega_0\sigma_z.
\end{eqnarray}

We study a single atom coupling to a single BEC and a two-component
BEC respectively.  We discuss these two cases in the following two
subsections.

\subsection{Case I: A BEC}
We first consider the single atom coupled to a condensate. The
Hamiltonian of a BEC confined in a trapping potential is given by,
\begin{eqnarray}
H_1&=&\int{dr}^3\Psi^\dag_1(r)\Big[-\frac{1}{2m_1}\nabla^2+V_1(r)\nonumber\\
&&+\frac{U_{11}}{2}\Psi^\dag_1(r)\Psi^\dag_1(r)\Psi_1(r)\Big]\Psi_1(r),
\end{eqnarray}
where $\Psi_1(r)$ is the annihilator field operator at the position
$r$, $V_1(r)$ is the external trapping potential, $U_{11}$ is the
self-interaction strength and $m_1$ is the mass of an atom.  The
condensate is assumed to be trapped in a deep potential such that
the BEC can be well described within the single-mode approximation
\cite{Milburn1,Dunningham}, i.e., $\Psi_1(r){\approx}a\psi_1(r)$.
Here $a$ and $\psi_1(r)$ are the annihilator operator and the mode
function of the condensate respectively.  Then, the Hamiltonian is
written as
\begin{eqnarray}
H_1&=&{\omega}_1a^\dag{a}+\kappa_1({a^\dag}a)^2,
\end{eqnarray}
where $\omega_1$ and $\kappa_1$ are the energy frequency and the
self-interaction strength respectively.

The single two-level atom interacts with the condensates via
coherent collisions.  The Hamiltonian describes such coupling as
\begin{eqnarray}
H^I_1&=&\kappa_{1e}|e\rangle\langle{e}|a^\dag{a},\\
&=&\kappa_{1e}(\sigma_z-1)a^\dag{a}.
\end{eqnarray}
where $\kappa_{1e}=2\pi{a_{1e}}\int{dr}^3|\psi_1^*(r)\psi_e(r)|^2/m$
and $\psi_e(r)$ is the wavefunction of the single atom and $a_{1e}$
is the s-wave scattering length between the atom at the upper state
$|e\rangle$ and the condensate. We consider this single atom trapped
in a state-dependent potential \cite{Treutlein} so that only the
upper state $|e\rangle$ interacts with the condensate
\cite{Bruderer}.  Besides, we assume that coherent collisions
between atom and the BEC will not further excite the motional state
of the single atom. Otherwise, it will give rise to the additional
noise and affect our detection scheme.

The size of the ground state wavefunction of the atom and the BEC
are roughly equal to the trap size. We denote the ``volume'' of the
trap as $V$. The interaction parameter $\kappa$ can be found as
$2{\pi}a_{1e}/mV$. In general, this interaction strength is weak.
Nevertheless, the scattering length $a_{1e}$ can be greatly
increased by tuning an external magnetic field near a Feshbach
resonance so that the interaction strength is greatly enhanced
\cite{Inoyue,Cornish,Marte,Roati}. This is a very useful tool for
controlling the atom-BEC coupling.

\subsection{Case II: Two-component BEC}
Next, we consider the single atom to be coupled to a two-component
BEC. The Hamiltonian of a two-component condensate is given by
\begin{eqnarray}
H_2&=&\int{dr}^3\Psi^\dag_{\alpha}(r)\Big[-\frac{1}{2m_\alpha}\nabla^2+V_\alpha(r)\nonumber\\
&&+\frac{U_{\alpha\alpha}}{2}\Psi^\dag_\alpha(r)\Psi_\alpha(r)
+\frac{U_{\alpha\beta}}{2}\Psi^\dag_\beta(r)\Psi_\beta(r)\Big]\Psi_\alpha(r),\nonumber\\
\end{eqnarray}
where $\Psi_\alpha(r)$ is the annihilator field operator for the
component $\alpha$, $V_\alpha(r)$ is the trapping potential,
$U_{\alpha\alpha}$ and $U_{\alpha\beta}$ are the intra-component and
inter-component interactions respectively, and $\alpha,\beta=1,2$.

As before, we adopt the single-mode approximation on the two
component condensates in which they are confined in a deep potential
\cite{Milburn1,Ng}.  We write
$\Psi_{\alpha}(r)\approx{\eta}\psi_{\alpha}(r)$, where $\eta=a,b$
and $\psi_\alpha(r)$ are the annihilation operator and the mode
function for the component $\alpha$ respectively.  Thus, the
Hamiltonian can be written as
\begin{equation}
H_2=\omega_1{a^\dag{a}}+\omega_2{b^\dag{b}}+\kappa_{1}(a^\dag{a})^2+\kappa_{2}(b^\dag{b})^2+\kappa_{12}a^\dag{a}b^\dag{b},
\end{equation}
where the energy frequency $\omega_\alpha$, the self-interaction
strength $\kappa_{\alpha}$ and the inter-component interaction
strength $\kappa_{12}$. It is convenient to express the Hamiltonian
in terms of the angular momentum operators as:
\begin{eqnarray}
H_2&=&\tilde{\omega}J_z+\tilde{\kappa}J^2_z,
\end{eqnarray}
where $J_x=(a^\dag{b}+b^\dag{a})/2$, $J_y=(a^\dag{b}-b^\dag{a})/2i$
and $J_z=(a^\dag{a}-b^\dag{b})/2$.  The parameters $\tilde{\omega}$
and $\tilde{\kappa}$ are $(\omega_1-\omega_2)/2$ and
$\kappa_1+\kappa_2-\kappa_{12}$ respectively.

We consider the single atom coupled to the two-component BEC via
coherent collisions.  The Hamiltonian for such an atom-BEC coupling
has the form:
\begin{equation}
H^I_2=|e\rangle\langle{e}|(\kappa_{1e}a^\dag{a}+\kappa_{2e}b^\dag{b}),
\end{equation}
where
$\kappa_{\alpha{e}}={2\pi{a_{\alpha{e}}}}\int{dr}^3|\psi^*_\alpha(r)\psi_e(r)|^2/m_\alpha$
and $a_{\alpha{e}}$ is the s-wave scattering length between the atom
in state $|e\rangle$ and the component $\alpha$. We can rewrite the
Hamiltonian in terms of the angular momentum operators as
\begin{equation}
H^I_2=(\kappa_{1e}-\kappa_{2e})(\sigma_z-1)J_z+(\kappa_{1e}-\kappa_{2e})N/2,
\end{equation}
where $N$ is the total number of atoms.  The constant term will be
omitted in the subsequent discussion. To strengthen the atom-BEC
coupling, we can increase the scattering length between the atom and
one of the components by adjusting a magnetic field approaching a
Feshbach resonance. For example, we can modulate the magnetic field
to increase the scattering length between the excited state
$|e\rangle$ and the component $\alpha=2$.

In fact, the collective excitations of the two mode condensates
behave like a harmonic oscillator.  This can be shown by taking the
leading approximation based on the Holstein-Primakoff transformation
(HPT) \cite{Holstein} to map the angular momentum operators into the
harmonic oscillators: $J_{+}\approx{\sqrt{N}c^\dag}$,
$J_{-}\approx{\sqrt{N}c}$ and $J_z=c^\dag{c}-N/2$.  This
approximation is valid as long as the excitations are very small
\cite{Ng}, i.e., $\langle{c^\dag{c}}\rangle/N{\ll}1$. Therefore, the
effective atom-BEC Hamiltonian can be readily obtained
\begin{eqnarray}
{\tilde{H}^I_2}&=&(\kappa_{1e}-\kappa_{2e})(\sigma_z-1)c^\dag{c}.
\end{eqnarray}
We assume the interaction strength $\kappa_{1e}$ and $\kappa_{2e}$
are unequal to each other, i.e., $\kappa_{1e}\neq\kappa_{2e}$. The
state of collective excitations of the BEC can be approximated as a
coherent state $|\alpha\rangle$ \cite{Barnett} and $|\alpha|^2$ is
the mean excitations of the two-component condensate.  The amount of
mean excitations can be adjusted by using a two-photon Rabi pulse
\cite{Myatt}.

\section{Scheme to Probe Decoherence}
As discussed above, both the single and two-component BEC's can be
described as harmonic oscillators. In other words, the single atom
(qubit) is effectively coupled to a harmonic oscillator.  In fact,
the Hamiltonian in both cases are of the same form and can be
written as:
\begin{eqnarray}
H&=&\omega_0\sigma_z+{\omega}d^\dag{d}+\kappa(d^\dag{d})^2+\chi({\sigma_z-1})d^\dag{d},
\end{eqnarray}
where $d$ and $\omega$ are the annihilation operator and the
frequency of the harmonic oscillator respectively,  $\kappa$ is the
strength of the nonlinearities and $\chi$ is the qubit-oscillator
coupling strength.

We consider the interaction strength $\chi$ is much stronger than
the strength $\kappa$ so that the effect of nonlinearities arising
from particle interactions can be ignored in the quantum dynamics.
The strength $\kappa$ is roughly equal to 100Hz \cite{Milburn1} for
the trap frequency around 1 kHz. The interaction strength $\chi$ can
be enhanced to several times $\kappa$ by approaching the Feshbach
resonance with a magnetic field. Therefore, the system is equivalent
to a qubit-harmonic oscillator system without any non-linearity.

We present a scheme to detect the decoherence of the BEC. The
dominant decoherence source of the BEC's is the atom loss due to
three-body inelastic collisions \cite{Burt}. The master equation for
the condensate is given by \cite{Jaksch,Jack1,Jack2}
\begin{equation}
\label{3body}
\dot{\rho}=\frac{\gamma_3}{6}[2d^3\rho{d^{\dag3}}-{d^{\dag3}}d^3\rho-\rho{d^{\dag3}}d^3],
\end{equation}
where $\gamma_3=K_3{\int}dr^3|\psi(r)|^6$, $K_3$ is the three-body
coefficient, $d$ and $\psi(r)$ is the destruction operator and the
mode function of the condensate mode respectively. Remarkably, in
the limit of large number of atoms, the master equation
($\ref{3body}$) can be well approximate to the one-body master
equation but with a new dissipation parameter $\Gamma$ \cite{Jack1}:
\begin{equation}
\label{1body}
\dot{\rho}={\Gamma}[2d\rho{d^\dag}-{d^\dag}d\rho-\rho{d^\dag}d],
\end{equation}
where ${\Gamma}=3N^2\gamma_3/2$.  For the case of two-component BEC,
we assume the atom loss mainly coming from one of the components,
say $\alpha=2$.   We note that losing one atom in the component
$\alpha=2$ results in one loss in the relative population
$\langle{J_z}\rangle$.  Thus, the process of atom loss in the
two-component BEC can be described by the master equation
($\ref{1body}$) in the large atom number limit.

The master equation (\ref{1body}) can be solved exactly and its
solution is best expressed in the coherent state basis. In our paper
we will require the time evolution of density operator terms of the
form $|\alpha\rangle\langle {\alpha}e^{i\theta}|$, where
$|\alpha\rangle$ and $|{\alpha}e^{i\theta}\rangle$ are two coherent
states differing by a rotation in phase space, under the master
equation (\ref{1body}). Up to a normalization constant, the time
evolution of the above term is of the form \cite{Walls,Davidovich},
\begin{eqnarray}
\tilde{\rho}(t)&\propto&|\alpha{e^{-{\Gamma}{t/2}}}\rangle\langle\alpha{e^{-{\Gamma}{t}/2}}|
+|\alpha{e^{i\theta-{\Gamma}{t/2}}}\rangle\langle\alpha{e^{i\theta-{\Gamma}{t}/2}}|\nonumber\\
&&+e^{-|\alpha|^2(1-e^{i\theta})(1-e^{-{\Gamma}{t}})}(|{\alpha}e^{-{\Gamma}{t}/2}\rangle\langle{\alpha}e^{i\theta-{\Gamma}{t}/2}|\nonumber\\
&&+e^{-|\alpha|^2(1-e^{-i\theta})(1-e^{-{\Gamma}{t}})}|{\alpha}e^{i\theta-{\Gamma}{t}/2}\rangle\langle{\alpha}e^{-{\Gamma}{t}/2}|),\nonumber\\
\end{eqnarray}
For the short times, i.e., $\Gamma{t}\ll{1}$, one can approximate
$(1-e^{{\Gamma}{t}})$ and $|\alpha{e^{-{\Gamma}{t/2}}}\rangle$ as
${\Gamma}{t}$ and $|\alpha\rangle$ respectively.  This is a good
approximation for an underdamped oscillator if the dectection
timescale $\chi^{-1}$ is much shorter than the dissipation timescale
${\Gamma}^{-1}$. Note that the decoherence time-scale
$({\Gamma|\alpha|^2})^{-1}$ can still be comparable to $\chi^{-1}$,
so that the decoherence can be detected.

In addition, we assume that the two-level atom has very long
coherence times so that it can act as a faithful microscopic probe
to detect the decoherence.  In fact, the long coherence times of the
atomic condensates with two hyperfine states have been measured
using Ramsey spectroscopy \cite{Harber}.

\subsection{Initial Coherent State} Our scheme is very simple in
which involves a few procedures only. First, we perform a unitary
transformation of the two-level atom to create an equal
superposition of the states $|g\rangle$ and $|e\rangle$ whereas the
BEC is prepared as a coherent state.  Such the unitary
transformation can be easily made by a Rabi pulse
\cite{Harber,Treutlein}. Initially, a separable state of the
two-level atom and the harmonic oscillator is considered as
\begin{eqnarray}
|\Psi(0)\rangle&=&|\psi\rangle_{Q}\otimes|\alpha\rangle,
\end{eqnarray}
where $|\psi\rangle_Q=(|e\rangle+|g\rangle)/\sqrt{2}$ and
$|\alpha\rangle$ is a coherent state. To manifest the evolution of
the atom-BEC system, we first consider the case without the
decoherence setting in. The atom becomes entangled with the harmonic
oscillator just after switching on the atom-BEC interaction. The
state can be written as
\begin{equation}
|\Psi(t)\rangle=\frac{1}{\sqrt{2}}(e^{-i\omega_0t}|e\rangle\otimes|\alpha(t)\rangle+e^{i\omega_0t}|g\rangle\otimes|e^{2i{\chi}t}\alpha(t)\rangle),
\end{equation}
where
$|{\alpha}(t)\rangle{\approx}e^{{-i}{\chi}{d^\dag{d}t}}|\alpha\rangle$,
for $\chi\gg\kappa$. The atom-BEC interaction gives rise to a
rotation of coherent state in phase space.  Thus, the phase shifts
of the condensate are acquired according to the states $|g\rangle$
and $|e\rangle$ respectively.  As a result, a superposition of two
coherent states are generated.

The ``distance'' $D(t)$ in phase space between the two states can be
defined as $2|\alpha|\sin\chi{t}$ \cite{Brune}.  The quantity $D$
can indicate the ``distance'' of the superpositions.  However, the
atom completely disentangles with the BEC at time $t'=\pi/\chi$ and
the state reads
\begin{eqnarray}
|\Psi(t')\rangle&=&\frac{1}{\sqrt{2}}(e^{i\omega_0{t'}}|g\rangle+e^{-i\omega_0{t'}}|e\rangle)\otimes|\alpha(t')\rangle.
\end{eqnarray}

Now we consider the system in the presence of the decoherence.  We
begin the investigation of the decoherence of the harmonic
oscillator in the underdamped case.  We must describe the system
with the density matrix because the system will evolve to a
statistical mixture. Initially, the density matrix is
\begin{eqnarray}
\rho_{\rm
c}(0)&=&|\psi\rangle\langle\psi|_Q\otimes|\alpha\rangle\langle\alpha|.
\end{eqnarray}
At time $t=t'/2$, the density matrix evolves as
\begin{eqnarray}
\rho_{\rm c}(t)&=&\frac{1}{2}[|g,\alpha_0\rangle\langle{g},\alpha_0|+|e,\alpha_1\rangle\langle{e},\alpha_1\rangle\langle{\alpha}_1|\nonumber\\
&&+e^{-\bar{\Gamma}/2}(e^{2i\omega_0{t}}|g,\alpha_0\rangle\langle{e},\alpha_1|+e^{-2i\omega_0{t}}|{e},\alpha_1\rangle\langle{g},\alpha_0|],\nonumber\\
&&
\end{eqnarray}
where $|g,\alpha_0\rangle=|g\rangle|\alpha_0\rangle$,
$|e,\alpha_1\rangle=|e\rangle|\alpha_1\rangle$, $|\alpha_0\rangle$
and $|\alpha_1\rangle$ are $|e^{2i\chi{t}}\alpha(t)\rangle$ and
$|\alpha(t)\rangle$ respectively.

The decoherence factor $e^{-\bar{\Gamma}/2}$ appears in the terms
$|g,\alpha_0\rangle\langle{e},\alpha_1|$ and
$|{e},\alpha_1\rangle\langle{g},\alpha_0|$. At the end of detection
times $t=t'$, the density matrix reads
\begin{eqnarray}
\rho_{\rm c}(t')&=&[|g\rangle\langle{g}|+|e\rangle\langle{e}|+e^{-\bar{\Gamma}}(e^{2i\omega_0{t'}}|g\rangle\langle{e}|+e^{-2i\omega_0{t'}}|e\rangle\langle{g}|]\nonumber\\
&&\otimes\frac{1}{2}|\alpha(t')\rangle\langle\alpha(t')|.
\end{eqnarray}
The atom disentangles with the condensate and the condensate is
brought back to its original state.

The atom-BEC coupling can be effectively switched off by tuning the
external magnetic field and then we can slowly move out the atom
from the BEC.   Then, the state of atom is measured and the
probability of single atom at the state $|e\rangle\langle{e}|$ can
be found as
\begin{eqnarray}
P(|e\rangle\langle{e}|)&=&\frac{1+e^{-\bar{\Gamma}}\cos2(\omega_0t'+\delta)}{2},
\end{eqnarray}
if the initial state of the atom
$|\psi(\delta)\rangle_Q=(|g\rangle+e^{i\delta}|e\rangle)/\sqrt{2}$
is considered, where $\delta$ is the phase shift between the states
$|g\rangle$ and $|e\rangle$. The partial coherence factor
$e^{-\bar{\Gamma}}$ appears in the probability of the state
$|e\rangle\langle{e}|$.  The measurement of the visibility of the
fringes as a function of $\delta$ leads us to determining the factor
$e^{-\bar{\Gamma}}$.

Since the instantaneous superposition of states decoheres as
${\Gamma}{D^2(t)}$, therefore the average value of the decoherence
rate $\langle{\Gamma}\rangle$ can be evaluated as
\begin{eqnarray}
\langle{\Gamma}\rangle&=&\frac{4{\Gamma}\chi|\alpha|^2}{\pi}\int^{\pi/\chi}_0d{t}\sin^2\chi{t}.
\end{eqnarray}
The average rate $\langle{\Gamma}\rangle$ can be obtained as
$2{\Gamma}|\alpha|^2$. Therefore, the decoherence factor
$\bar{\Gamma}=\langle\Gamma\rangle\pi/\chi$ after the completion of
the probing process is given by $2\pi{\Gamma}|\alpha|^2/\chi$. We
can probe the decoherence  factor $\bar{\Gamma}$ by measuring of the
probability of the excited state of the single atom.  Hence, the
damping rate $\Gamma$ of the BEC can be determined.

\subsection{Initial Thermal State}
The temperature of the BEC is nearly absolute zero, indeed its
temperature ranges from 100 nK to 500 pK \cite{Bradley,Leanhardt}.
The state of the BEC can be well described as a thermal state if the
finite temperature is taken account. Our detection can be used for
probing the decoherence with the initial thermal state. This enables
us to study the decoherence due to the finite temperature effect.
Initially, the density matrix is of the form
\begin{eqnarray}
\rho_{\rm
th}(0)&=&|\psi\rangle\langle\psi|_Q{\otimes}{\int}d^2{\alpha}p(\alpha)|\alpha\rangle\langle\alpha|,
\end{eqnarray}
where $p(\alpha)$ are probabilities
$\exp(-|\alpha|^2/\bar{n})/\pi\bar{n}$.  The evolution of the
density matrix at time $t=t'/2$ is
\begin{eqnarray}
&&\rho_{\rm th}(t)\nonumber\\
&=&{\int}d^2{\alpha}\frac{p(\alpha)}{2}[|g,\alpha_0\rangle\langle{g},\alpha_0|+|e,\alpha_1\rangle\langle{e},\alpha_1|+e^{{-\bar{\Gamma}'_\alpha/2}}\nonumber\\
&&\times(e^{2i\omega_0{t}}|g,\alpha_0\rangle\langle{e},\alpha_1|+e^{-2i\omega_0{t}}|e,\alpha_1\rangle\langle{g},\alpha_0|)],\nonumber\\
&&
\end{eqnarray}
where $e^{-\bar{\Gamma}'_\alpha}$ is the decoherence factor at time
$t$ for each $\alpha$. Finally, the atom disentangles with the
condensate and the density matrix is found to be
\begin{eqnarray}
\rho_{\rm
th}(t')&=&\frac{1}{2}[|g\rangle\langle{g}|+|e\rangle\langle{e}|+e^{-\bar{\Gamma}'}(e^{2i\omega_0t'}|g\rangle\langle{e}|\nonumber\\
&&+e^{-2i\omega_0t'}|e\rangle\langle{g}|)]\otimes{\int}d^2{\alpha}{p(\alpha)}|\alpha(t')\rangle\langle\alpha(t')|\nonumber\\
&& \label{mixed}
\end{eqnarray}
The decohernce factor
$e^{-\bar{\Gamma}'}={\int}d^2{\alpha}p(\alpha)e^{\bar{\Gamma}'_\alpha}$
sums up all contributions from the decoherence of the different
possible coherent states $|\alpha\rangle$.  Similarly, the
decoherence factor $e^{-\bar{\Gamma}'}$ can be detected through the
measurement of the visibility of the atom. We have shown that our
scheme can be used to probe the decoherence of the condensates with
the initial coherent and thermal states. In fact, as
Eq.(\ref{mixed}) is valid for any distribution $p(\alpha)$, the
scheme is valid for any mixture of coherent states being the initial
state of the condensate. So for example, if the amplitude of the
coherent state was known but its phase was completely unknown, our
method of probing decoherence would still be applicable.

\section{Decoherence Sources}
The main decoherence sources of the BEC are three-body inelastic
collisions and collisions with the background gas \cite{Burt}.  We
sort out several noise sources of the BEC's and discuss them as
follows:

\subsection{Atom loss}
{\it Background gases and spontaneous light scattering:}  The
background gases and spontaneous emission contribute the one-body
loss and thus induce the decoherence of the BEC's \cite{Stenger}.
The loss rate $\Gamma_1$ is of the form $K_1{N}$, where $K_1$ is the
one-body loss coefficient.  However, such decoherence effect is weak
in the current experiment circumstance \cite{Stenger}.

{\it Two-body and three-body inelastic collisions:} The two-body
inelastic collisions are very rare in the atomic condensates and its
loss rate is $K_2N^2/V$, where $K_2$ is the two-body coefficient and
$V$ is the volume of the trapped BEC. The inelastic collisions
mainly occurs due to the three-body collisions
\cite{Burt,Marte,Stenger}. The rate of decoherence $\Gamma$ is
$K_3N^3/V^2$, where $K_3$ is the three-body coefficient.  The
three-body coefficient $K_3$ is about ${\sim}10^{-29}$ ${\rm
cm}^6{\rm s}^{-1}$ \cite{Burt,Marte}. It is noted that inelastic
collisions are greatly enhanced near Feshbach resonances
\cite{Stenger}. However, we assume this effect is minimal to our
detection scheme if the probing times are very short.

\subsection{Phase damping}
Phase damping describes a process of the loss of the coherence
without losing energy. Elastic collisions between the BEC and the
surrounding gases can cause phase damping
\cite{Anglin,Vardi,Anglin2}. The elastic scattering with vacuum
noises can also contribute the dephasing.  However, our scheme is
not applicable in detecting the phase damping for the coherent
states with the same magnitude $|\alpha|^2$ but with the different
phases. The decoherence factor cannot be imprinted on the atom. In
fact, the dephasing rate $\Gamma_p$ depends on the temperature of
the gases and therefore its rate is very low in the current
experiments of the BEC \cite{Anglin,Vardi,Anglin2}.  The decoherence
timescale of the phase damping is much longer than that of the atom
loss. Thus, the dephasing effect is negligible compared to the atom
loss.

\section{Discussion}
Our scheme involves the active control of the scattering length
using a magnetic field approaching Feshbach resonances. We can
greatly increase the scattering length $a$ with the Feshbach
resonance and therefore speed up the process of creating
superpositions. The scattering length $a$ is varied as a function of
external magnetic field B \cite{Stenger,Donley}:
\begin{eqnarray}
a&=&a_{bg}\Big(1+\frac{\Delta}{B_{0}-B}\Big),
\end{eqnarray}
where $a_{bg}$ is the off-resonant scattering length, $B_0$ is the
resonant magnetic field and $\Delta$ is the width of the Feshbach
resonance.

Nevertheless, the change of scattering length $a$ will accompany
with an increasing three-body inelastic collisions rate. The rate of
three-body collisions loss increases to 20(60) times for the low
magnetic field of sodium atoms \cite{Stenger}.  This limits the use
of Feshbach resonance to create a superposition state.  In the worst
case, the single atom may be lost due to the formation of molecules
\cite{Donley}.  Then, our scheme is no longer applicable -- of
course, if we loose the atom from our trap, we no longer continue
with that run of the experiment and simply restart the experiment
with a fresh atomic qubit. It is quite possible, though, to maintain
the coherence of the single atom and create superpositions with a
fast ramp speed \cite{Marte}. Thus, our scheme can be applied to the
situation that the detection rate is much shorter than the decay
rate.

\section{Conclusion}
In summary, we have studied the coupling of a single atom to the
single and two-component BEC's respectively.  We have presented a
scheme to create mesoscopic superpositions involving distinct
classical-like (or coherent) states of the BEC and probe their
decoherence. The probing of the decoherence is applicable to both
initial coherent and thermal states of the BEC. Only the state of
the atom state needs to be directly measured in this experiment to
probe the decoherence of the BEC. The various noise sources leading
to the decoherence of the condensates are also discussed. This
allows us to investigate the major decoherence source due to atom
loss in detail.

\begin{acknowledgments}
The work of H. T. Ng is supported by the Quantum Information
Processing IRC (QIPIRC) (GR/S82176/01). S. Bose also thanks the
Engineering and Physical Sciences Research Council (EPSRC) UK for an
Advanced Research Fellowship and the Royal Society and the Wolfson
Foundation.

\end{acknowledgments}

\end{document}